\documentclass[preprint,superscriptaddress,amsmath,amssymb,prl,floatfix,a4paper,onecolumn]{revtex4-1} 
\usepackage[ngerman,english]{babel}
\usepackage[utf8]{inputenc}   %Umlaute
\usepackage{xspace}
\usepackage{amsmath,tensor} 
\usepackage{array,color}
\usepackage{relsize,exscale}  %\{relsize{-2} or \normalsize
\usepackage{textcase}	% \MakeTextUppercase{adlkjf $3+4$}
\usepackage{soul,color}		%\so{} or \hl{} or \ul{} or \st{} or \caps{}
\usepackage{calc} %for boxes
\usepackage{makeidx,showidx}
\usepackage{graphicx}
\usepackage{wrapfig}
\usepackage{float}
\usepackage{textcomp}
\usepackage{siunitx}
\usepackage[version=3]{mhchem}
\usepackage{enumitem}
\usepackage{fancyref}
\usepackage[hidelinks=true]{hyperref}
\sethlcolor{green}
\setulcolor{red}
\usepackage[running]{lineno}			%\linenumbers
\usepackage{xspace}

\newcommand{\erww} [1] {\ensuremath{\langle {#1} \rangle}}

\newcommand{\lsco} {{La$_{2-x}$Sr$_x$CuO$_4$}\@\xspace}

\newcommand{\ybcoF} {$\ce{YBa2Cu3O_{7}}$\@\xspace}
\newcommand{\ybco} {$\ce{YBa2Cu3O_{6+y}}$\@\xspace}

\newcommand{\ybcoE} {$\ce{YBa2Cu4O8}$\@\xspace}

\newcommand{\beq} {\begin{equation}}
\newcommand{\eeq} {\end{equation}}

\begin{document}
\title{Stripe-like correlations in the cuprates from oxygen NMR}
\author{Daniel Bandur}
\author{Abigail Lee}
\author{Stefan Tsankov}
\affiliation{University of Leipzig, Felix Bloch Institute for Solid State Physics, 04103 Leipzig, Germany}
\author{Andreas Erb}
\affiliation{Walther Meissner Institut, Bayerische Akademie der Wissenschaften, 85748 Garching, Germany}
\author{Jürgen Haase$^{*}$}
\affiliation{University of Leipzig, Felix Bloch Institute for Solid State Physics, 04103 Leipzig, Germany}\par\medskip

%\affiliation[2]{Walther Meissner Institut, Bayerische Akademie der Wissenschaften, 85748 Garching, Germany} 
\vspace{1cm}
\begin{abstract}
{\footnotesize Nuclear magnetic resonance (NMR) of planar oxygen, with its family independent phenomenology, is ideally suited to probe the nature of the quantum matter of superconducting cuprates. Here, with new experiments on La$_{2-x}$Sr$_x$CuO$_4$, in particular also at high doping levels, we report on short-range stripe-like correlations between local charge and spin. Their amplitudes at room temperature are nearly independent of doping up to at least $x=0.30$, only their relative phase slips near $x=1/4$. Comparisons show the correlations to be generic to the cuprates. Despite the atomic scale length, the variations still resemble the average spin and charge relation, which is not expected from the otherwise simple, apparently metallic behavior, even far into the overdoped regime. Perhaps the phase slip is at the heart of a quantum critical point that demands pseudogap behavior towards lower doping levels in an otherwise strange metal.}
\end{abstract}

\maketitle
% \linenumbers

\section{Introduction}
When  high temperature superconductivity in the cuprates was discovered \cite{Bednorz1986} it was perhaps not expected that its understanding would still be controversial today. The likely special nature of the quantum matter governing the electronic properties of the CuO$_2$ plane was pointed out early on \cite{Zaanen1989,Emery1993,Cheong1991,Tranquada1995,Bianconi1996}. Over the years, experimental and theoretical evidence proliferated, but even today, among other things, stripes remain a topic of great discussion \cite{Tranquada2020}, as well as intra-unit-cell modulation \cite{Fujita2014}. Meanwhile, recent experimental results have called into question whether the overdoped materials can be described as Fermi liquids \cite{VanHeumen2022, Ayres2021}, and first principles calculations find competing charge and magnetic orders \cite{Zhang2020}, including well into the overdoped regime \cite{Jin2024}.  

Nuclear magnetic resonance (NMR) as a bulk, local probe must carry the properties of the electronic matter, but since there is no microscopic theory involving NMR, it is difficult to conclude on the meaning of some of the NMR findings. For example, the still controversial pseudogap was readily discovered by NMR shifts that already showed a temperature dependence far above the critical temperature $T_\mathrm{c}$ \cite{Alloul1989} at lower doping levels, while an explanation is still missing.

\begin{figure}[t]
\centering
\includegraphics[width=.6\textwidth]{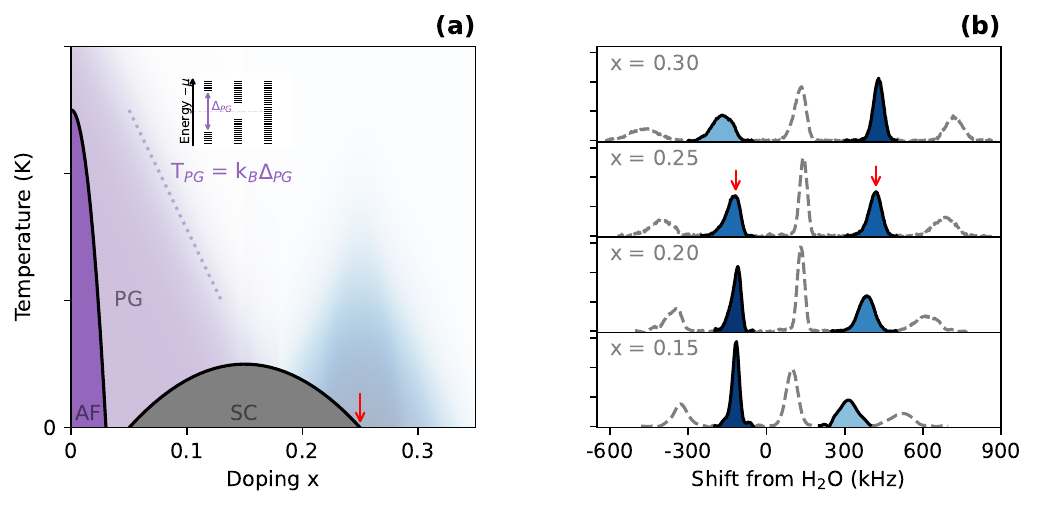}
\caption{(a) Main panel: Common cuprate phase diagram as a function of doping, $x$, with the pseudogap line as determined from NMR \cite{Bandur2023}. Inset: A gap $\Delta_\mathrm{PG}$ in a universal density of sates describes the high-temperature NMR for planar O and its values were used for the main panel. (b) $^{17}$O NMR spectra of La$_{2-x}$Sr$_x$CuO$_4$ at 300 K reveal a doping dependent critical flip in the spectral symmetry at $x=0.25$, and a doping independent resonance frequency of the first lower satellite below $x=0.25$. It will be argued that this is due to a stripe-like correlation that undergoes a critical change at the doping level $x=1/4$.}
\label{fig:fig1}
\end{figure}

Here, as will be explained in the main text further below, we report with new $^{17}$O NMR of \lsco on a critical crossover at $x=0.25$ of correlations between charge and spin, which reach across the dome, far into the overdoped regime, cf.~Fig.~\ref{fig:fig1}(a). Our results show that even the strongly overdoped \lsco cannot be a simple Fermi liquid (as suggested by others \cite{Li2023}). The main evidence comes from spectra  summarized in Fig.~\ref{fig:fig1}(b) that undergo a fundamental change with doping. We then make the case, by comparing to literature data, that the detected behavior is quite universal and describes the planar matter, and that it must describe stripe-like correlations with a phase change at high doping levels. \par\medskip

%In order to understand the significance of these findings that involve correlated spin and charge variations, one needs to be reminded of other cuprate NMR, including the more recent findings, which were already shown to present a different view of cuprate properties \cite{Bandur2023}. 

So far, cuprate NMR has mainly been discussed with a reference to the NMR spin shift $(K_\mathrm{S})$ and Heitler-Teller relaxation $(1/T_1)$ of nuclei, as both parameters relate to the electronic spin through magnetic hyperfine interaction \cite{Knight1949, Heitler1936}. Famously, the Korringa relation between both properties holds for metals, independent of the hyperfine scenario \cite{Korringa1950}. Early NMR for superconducting cuprates found that spin shift and magnetic relaxation do not follow simple metallic behaviour, but below $T_\mathrm{c}$, one finds at least qualitatively what is expected from a spin-singlet superconductor (magnetic shift and relaxation disappear rather rapidly as the temperature is lowered \cite{Hebel1957,Yosida1958}). 

Less well appreciated is the fact that nuclei with larger spin quantum numbers ($I>1/2$) also measure  electric field gradients $V_{ij}$ at the nucleus through electric hyperfine interaction ($^{17}$O has $I=5/2$, $^{63,65}$Cu have $I=3/2$). It leads to the well-known quadrupolar splitting $(\nu_Q)$ of the NMR resonance lines and can  cause nuclear relaxation \cite{Pound1950}. The splitting in zero or high magnetic fields can give direct evidence of any charge density variation \cite{Follstaedt1976}. 

In terms of the electric quadrupole interaction of planar Cu and O, one obviously must find strong doping dependences since the hole content of the CuO$_2$ plane changes tremendously with doping. How this relates to other observable properties is not clear and the focus was on fixed-doping temperature dependences. On the other hand, $^{17}$O NMR found very unusual correlated shifts and splittings \cite{Haase2000,Haase2002} that are not understood in a deeper way. It was also argued that the splitting of $^{17}$O quadrupole lines in YBa$_2$Cu$_3$O$_{6+y}$ could be caused by rather commensurate charge density variations \cite{Haase2003}, and not just by the materials orthorhombicity. This was only recently proven to be correct by NMR \cite{Reichardt2018}. Certainly, it must be related to the intra-unit-cell charge variation that was meanwhile discovered by STM \cite{Fujita2014}. However, inhomogeneity from chemical doping and sample quality inherently interfere with the characterization of intrinsic charge variations. At least, uninteresting doping variations are difficult to discern from intrinsic effects \cite{Singer2005}.\par\medskip

In terms of the spin shift, it appeared originally that planar Cu and O show similar temperature dependences, which is expected in simple scenarios for a uniform response to a magnetic field \cite{Takigawa1991,Bankay1994}. Today, however, it is certain, from the analysis of all available NMR literature data, that this is not the case \cite{Avramovska2022}. However, the data reveal that the magnetic planar $^{17}$O NMR is quite independent of family and  can be understood with a temperature independent pseudogap that decreases with doping and is located in a rather universal DOS \cite{Nachtigal2020}. These more recent findings agree with electronic entropy  \cite{Loram1998}. Planar Cu, however, shows a distinct family dependence. As one might expect, the planar Cu shift is the sum of a term that is also observed for planar O, and an additional term that is not experienced by planar O \cite{Bandur2023}. Therefore, the lack of complexity makes planar $^{17}$O NMR an ideal tool to study the response of the CuO$_2$ plane.\par\medskip

Importantly, and only developed in recent years, the charges at planar Cu and O can be measured with NMR. This gave remarkable insight. For example, it was found that different cuprate families share the nominal hole content differently between planar Cu and O. Indeed,  the hole content at planar O at optimal doping is proportional to the maximum critical temperature $T_\mathrm{c}$ \cite{Jurkutat2014,Rybicki2016}. This unexpected finding was confirmed with cellular DMFT very recently \cite{Kowalski2021}. This scenario also explains why pressure can increase $T_\mathrm{c}$ beyond what can be achieved by doping \cite{Jurkutat2023}.

Thus, planar $^{17}$O NMR offers a material independent view into the cuprate properties. In addition, planar $^{17}$O NMR shifts are rather reliable since orbital shift contributions are small \cite{Renold2003}, in contrast to the uncertainty at planar Cu.

%%%%%%%%%%%%%%%%%%%%%%%%%%%%%%%%
%%%%%%%%%%%%%%%%%%%%%%%%%%%%%%%%
 \section{Experimental}
%%%%%%%%%%%%%%%%%%%%%%%%%%%%%%%%
We investigated single crystals of La$_{2-x}$Sr$_x$CuO$_4$ with x=0.15, 0.20, 0.25, 0.30 grown via the travelling solvent floating zone (TSFZ) method. The starting materials were of a purity of at least 99.99 at.\%. The TSFZ method is a container-less crystal growth method where no crucible is involved, thus no impurities can be introduced from a container. Furthermore, the growth via TSFZ ensures a homogenous doping (Sr) distribution in the grown crystals and an overall purity which is the same as the starting materials or even better due the intrinsic refinement of this method \cite{Erb2015}. The growth crystals were oriented using the X-ray Laue Backscattering method with a accuracy of better than 0.5 degrees  and cut into the desired sample shape.  

While single crystals have weak sensitivity at low temperatures, aligned powders, often used previously, suffer from grain misalignment and thus from intrinsic, alignment-based correlations between shift and splitting. All single crystals are roughly $1$mm$\times 1$mm $\times 1$mm in size and are cut parallel to the c-axis. The single crystals  were exchanged under $^{17}$O atmosphere for 3 days at $\sim 850 ^{\circ}$C. From intensity measurements, the exchange is estimated to be $40\%$ to $90 \%$ (estimated from the planar Cu resonance) for the different samples. Samples showed the expected $T_\mathrm{c}$, based on the monitored change in RF coil inductance. All room temperature experiments have been carried out in a static magnetic field $B_0 = 11.74$T. Additional temperature- and field dependent measurements were performed in a variable field magnet (max.~$16$T). The presented resonances were recorded with Hahn echoes and typical pulse lengths of  $2\mu$s and separation times of $30\mu$s. Apical oxygen signals ($T_1 \approx 3$s) have been suppressed by use of fast repetition times ($\sim30$ms). The RF pulse power was adjusted for each transition. Planar oxygen shifts are referenced to regular tap water (higher order contributions in the quadrupole term can be neglected). Linewidths, if not stated otherwise, are given in half width at half maximum (HWHM).

%%%%%%%%%%%%%%%%%%%%%%%%%%%%%%%%
 \section{Correlated Linewidth Analysis}
%%%%%%%%%%%%%%%%%%%%%%%%%%%%%%%%

Planar $^{17}$O NMR shift (and relaxation) measurements are available in the literature and relate to a simple phenomenology of a doping dependent, but temperature independent pseudogap \cite{Nachtigal2020}. Missing to some extent are data at high doping levels, and in particular dedicated studies of the total linewidth as shown in Fig.~\ref{fig:fig1}(b).

In order to understand the lines in Fig.~\ref{fig:fig1}(b), we note that each of the 5 lines corresponds to a resonance between a pair of the 6 nuclear Zeeman states ($m$) with $\Delta m =1$. The detailed energies $E_m$ are influenced by internuclear or magnetic and electric hyperfine interactions of the electrons with the nuclear spin. In fact, we know from spin echo decay experiments that the NMR linewidths are inhomogeneous to a good approximation, i.e. each nuclear level, $E_m$, is very well defined on the time scale of the inverse linewidth that we discuss here. The nuclear Hamiltonian, for a particular alignment $\alpha$ of the crystal with respect to the magnetic field, can then be written as a sum of a magnetic and a quadrupolar term,
\begin{equation}\label{eq:ham}
\mathcal{H} = \hbar (1+K_\mathrm{\alpha}){\omega _\mathrm{ref}}{I_z} + \frac{{3I_z^2 - I(I + 1)}}{{\hbar 4I(2I - 1)}} eQ\cdot{V_{ZZ,\alpha}}
\end{equation}
where $K$ is the total magnetic shift. $\nu_\mathrm{ref}=\omega_\mathrm{ref}/2\pi$ is the resonance frequency of $^{17}$O in tap water, and the quadrupole splitting frequency is,
\begin{equation}\label{eq:eggs}
\nu_\mathrm{Q}=\omega_\mathrm{Q}/2\pi  = \frac{{3eQ}}{{2I(2I - 1)}h} \cdot{V_\mathrm{ZZ}},
\end{equation}
with the nuclear spin $I$, quadrupole moment $eQ$, and the electric field gradient's largest principle axis component $V_\mathrm{ZZ}$, which for planar O is along the $2p_\sigma$ bond. From the magnetic terms (sum of spin shift, orbital shift, dipolar and indirect coupling), the spin shift $K_\mathrm{S}$ dominates (anisotropic linewidths for various materials show the indirect coupling to be less than 10 kHz).

%%%%%%%%%%%%%%%%%%%%%%%%%%%%%%%%
\begin{figure}[t!]
\centering
\includegraphics[width=0.7\textwidth]{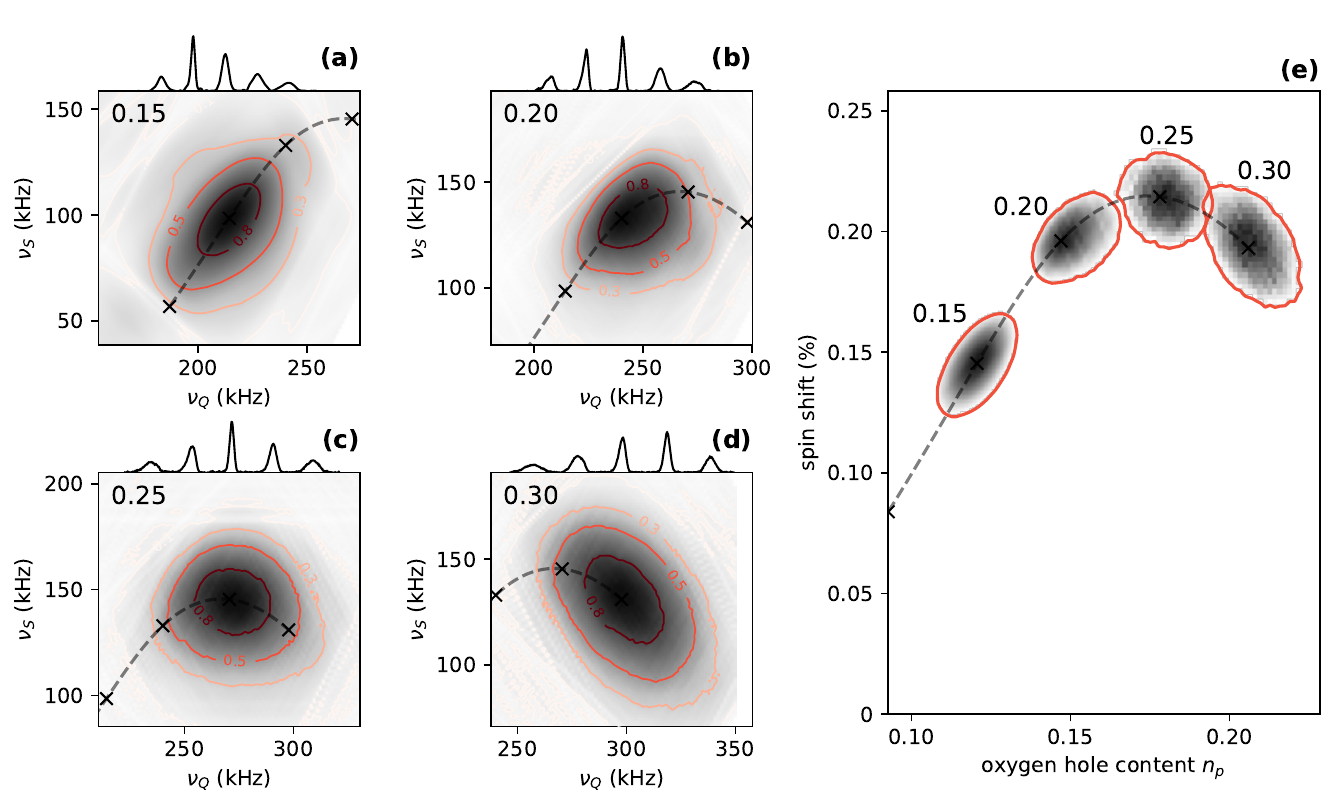}
\caption{(a) to (d): Analysis of the room temperature $^{17}$O NMR lineshapes of \lsco for doping levels of (a) $x= 0.15$, (b) $0.20$, (c) $0.25$, and (d) $0.30$. On top of each panel is the total lineshape from Fig.~\ref{fig:fig1}. Contours and the color gradient are the normalised probability $P$ of the pair $(\nu_{Q}, \nu_{S})$ contained within the measured lineshape, according to \eqref{eq:freq0}: $P(\nu_{Q}, \nu_{S}) = \prod_n \left(I(\nu_{S}+n\nu_{Q})\right)^{1/5}$, $I(\nu)$ is the signal intensity at a chosen frequency $\nu$ out of $\nu_\mathrm{n}$ in \eqref{eq:freq0}. The dashed lines connect the average shifts measured independently as a function of doping and thus $n_\mathrm{p}$. (e) Summary plot of all doping levels from (a) to (d), with the abscissa converted to the total planar O hole content, $n_\mathrm{p}$.}
\label{fig:fig2}
\end{figure}
%%%%%%%%%%%%%%%%%%%%%%%%%%%%%%%%
We will write the spin shift in frequency units, $\nu_\mathrm{S}= K_\mathrm{S}\cdot\nu_\mathrm{ref}$. This notation is more convenient here since the quadrupole splitting $\nu_\mathrm{Q}$ does not depend on the magnetic field. 
%For the cuprates, it is clear that $\nu_\mathrm{Q}$ changes linearly with the planar O hole content $n_\mathrm{p}$ of the $2p_\sigma$ orbital \cite{Haase2004} (see also further below).

From the Hamiltonian \eqref{eq:ham} one finds the following 5 resonance frequencies for a particular nucleus $j$,
\begin{equation}\label{eq:freq0}
\nu_\mathrm{n}^j = \nu_\mathrm{S}^j + n\cdot\nu_\mathrm{Q}^j \qquad  n = -2, -1, 0, 1, 2,
\end{equation}
where we note that the central line ($n=0$) is not affected by quadrupole interaction (no higher order terms necessary) and measures only the magnetic $\nu_\mathrm{S}^j$. We added the index $j$ to underline that each nucleus contributes independently to all 5 lines according to \eqref{eq:freq0}. In other words, a particular frequency slice of the central transition, $\Delta \nu_\mathrm{S}$, is proportional to the number of planar O nuclei that resonate in this frequency window, while the corresponding quadrupolar splittings of the contributing nuclei could be substantially different.
%In this sense we have atomic scale resolution for measuring charge and spin. Note, however, that one might expect that the electron spin density is distributed over a number nuclei. 

Given \eqref{eq:freq0}, the distribution of magnetic shift (inherent in all transitions) is just the central lineshape, but the satellites prohibit a simple estimate of the distribution of $\nu_\mathrm{Q}$ if shift and splitting are correlated. 

In an unbiased analysis, we take a particular frequency slice of the central line, $\Delta\nu_\mathrm{S}$, and calculate the probability that its nuclei contribute to a particular quadrupole splitting, $\Delta \nu_\mathrm{Q}$. This means, they have to contribute according to \eqref{eq:freq0} to all four satellites. This probability, $P(\nu_\mathrm{S}, \nu_\mathrm{Q})$, is taken as the product of the corresponding spectral intensities for all 5 transitions. The result is shown in Fig \ref{fig:fig2} in terms of contour plots.
As expected, the most probable pair ($\nu_\mathrm{S}, \nu_\mathrm{Q}$) is close to the main axis of an ellipse in Fig.~\ref{fig:fig2}, and tracks the average shift, $\erww{\nu_\mathrm{S}}$, and splitting, $\erww{\nu_\mathrm{Q}}$, that one can estimate directly from the center of gravity of the spectra. This is seen more easily in Fig.~\ref{fig:fig2}(e).

We note in Fig.~\ref{fig:fig2} that the doping dependence of the widths is rather weak. In other words, the magnetic and charge distributions are rather similar in size and doping independent.

\begin{figure}
\centering
\includegraphics[width=0.6\textwidth]{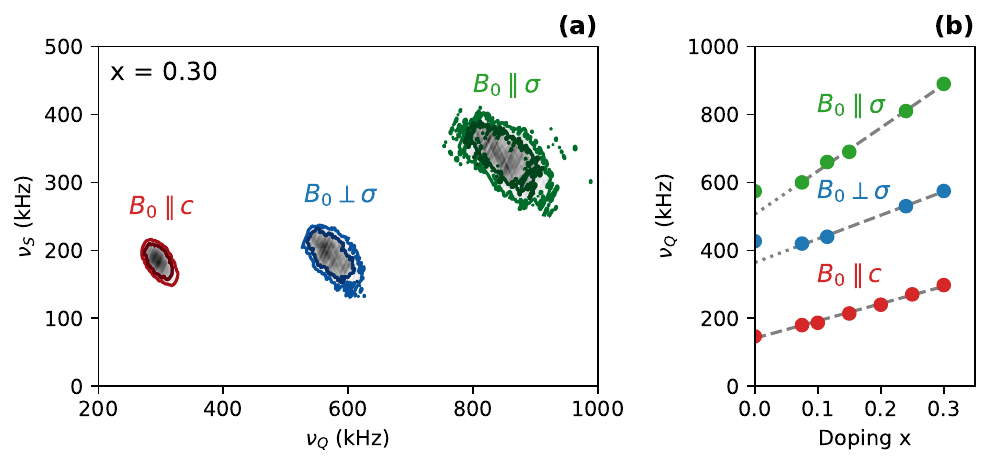}
\caption{(a) The probability contours  for $x=0.30$ and three directions of the magnetic field $B_0$ in terms of splitting and shift (same scale on both axes) for the $^{17}$O NMR at $11.74$ T (description in the main text). The data prove that it is the axial hole in $2p_\sigma$ that is varied, not the total electric field gradient that partially arises from the lattice. A variation of the isotropic spin response leads to the observed widths in $\nu_\mathrm{S}$ due to the anisotropic hyperfine coefficient \cite{Huesser2000} (see main text). (b) The doping dependence of the average splittings for the different field alignments \cite{Haase2004, Zheng1993, Imai2018, Walstedt2001}.}\label{fig:fig3}
\end{figure}

Given the relation between $\erww{\nu_\mathrm{S}}$ and $\erww{\nu_\mathrm{Q}}$, the dashed lines in Fig.~\ref{fig:fig2}, one may expect straight line segments along the dashed lines from a distribution, instead of ellipses with the main axis along the mean line. Circles as for $x=0.25$ are not expected at all since they document that the magnetic broadening is no longer correlated with the charge broadening, while the size of the uncorrelated part (17 kHz) of the magnetic broadening has not changed. At $x=0.30$ the main axis of the ellipse flips to follow the average relation again, now inverted with respect to $\erww{\nu_\mathrm{S}}$. Before we discuss the findings further, we turn to Fig.~\ref{fig:fig3} for an important check of consistency.

%%%%%%%%%%%%%%%%%%%%%%%%%%%%%%%%
%%%%%%%%%%%%%%%%%%%%%%%%%%%%%%%%

In Fig.~\ref{fig:fig3} the contour plots for $x=0.30$ and 3 different oxygen sites with respect to the magnetic field $B_0$ are shown: $B_0\parallel c$, i.e.\@ field parallel to the crystal $c$-axis (both O atoms in the CuO$_2$ unit cell are not distinguishable); $B_0\parallel \sigma$, field in the plane and along the $2p_\sigma$ bond; $B_0 \perp \sigma$, field in the plane and perpendicular to the $\sigma$ bond. The magnetic shift is related to the spin polarization by $\nu_\mathrm{S} = C_\alpha \erww{S_z}$, with the well-known anisotropic magnetic hyperfine coefficient $C_\alpha$ \cite{Renold2003} and the isotropic spin polarization along the quantization axis. $C_\alpha$ along $p_\sigma$  is about twice as large if $B_0\parallel \sigma$ \cite{Huesser2000}. This is what we indeed observe for both the magnetic shift and the magnetic width. 

Very different are the results for splittings and distributions in Fig.~\ref{fig:fig3}. There we find a mean splitting ratio of  [1 : 2 : 3], while from the distribution we have approximately [1 : 1 : 2]. At first glance, this may be surprising, but it is exactly what we expect if only the hole content in the 2$p_\sigma$ orbital is modulated, rather than the total electric field gradient (which is asymmetric, different from a hole content). It is well known that doping the cuprates leads to changes in the Cu and O hole contents that add to a background electric field gradient given by the chemistry \cite{Jurkutat2014}. In fact, the change in the planar O splitting with doping measures the $2p_\sigma$ hole content $n_\mathrm{p}$ quantitatively \cite{Haase2004}, so  $\Delta \nu_\mathrm{Q} \propto \Delta n_\mathrm{p}$.
Thus, the results in Fig.~\ref{fig:fig3} prove that, as discussed in Fig.~\ref{fig:fig2}, it is $n_\mathrm{p}$ and the isotropic electronic spin response $\erww{S_z}$ that are modulated.

Thus, the contour plots in Fig.~\ref{fig:fig3} reveal that one faces a planar hole variation of about $\Delta n_\mathrm{p} = 0.029(3)$, similar to what is found at lower doping levels. The magnetic broadening is still correlated with this hole variation, only the sign of the correlation has changed compared to lower doping levels. 

%%%%%%%%%%%%%%%%%%%%%%%%%%%%%%%%%%%%%%%%%%%%%
%%%%%%%%%%%%%%%%%%%%%%%%%%%%%%%%%%%%%%%%%%%%%%%%
\section{Discussion}
%%%%%%%%%%%%%%%%%%%%%%%%%%%%%%%%%%%%%%%%%%%%%%%%%%
%%%%%%%%%%%%%%%%%%%%%%%%%%%%%%%%%%%%%%%%%%%%%%%%

NMR is essentially a \emph{static} probe when it measures shifts and splittings. Fluctuating spin and charge predominantly contribute to nuclear relaxation. Indeed, accounts for both sorts of fluctuations exist for the cuprates. Heitler-Teller relaxation  persists for planar O over a large range of temperature and doping; only the states lost from the pseudogap lead to a defined decrease in the rate \cite{Nachtigal2020}. Near $T_\mathrm{c}$, quadrupolar relaxation was reported as well \cite{Suter2000}, but not studied in detail for other systems. Very slow fluctuations that may not be averaged fully during the nuclear precession, likely lead to strong relaxation or rapid spin echo decay, neither of which is observed here. Critical slowing down leads to signal wipe-out in underdoped materials \cite{Hunt2001}.

The average splittings and shifts have been studied over the years, and we give only a short summary here. 

\begin{figure}
\centering
\includegraphics[width=0.8\textwidth]{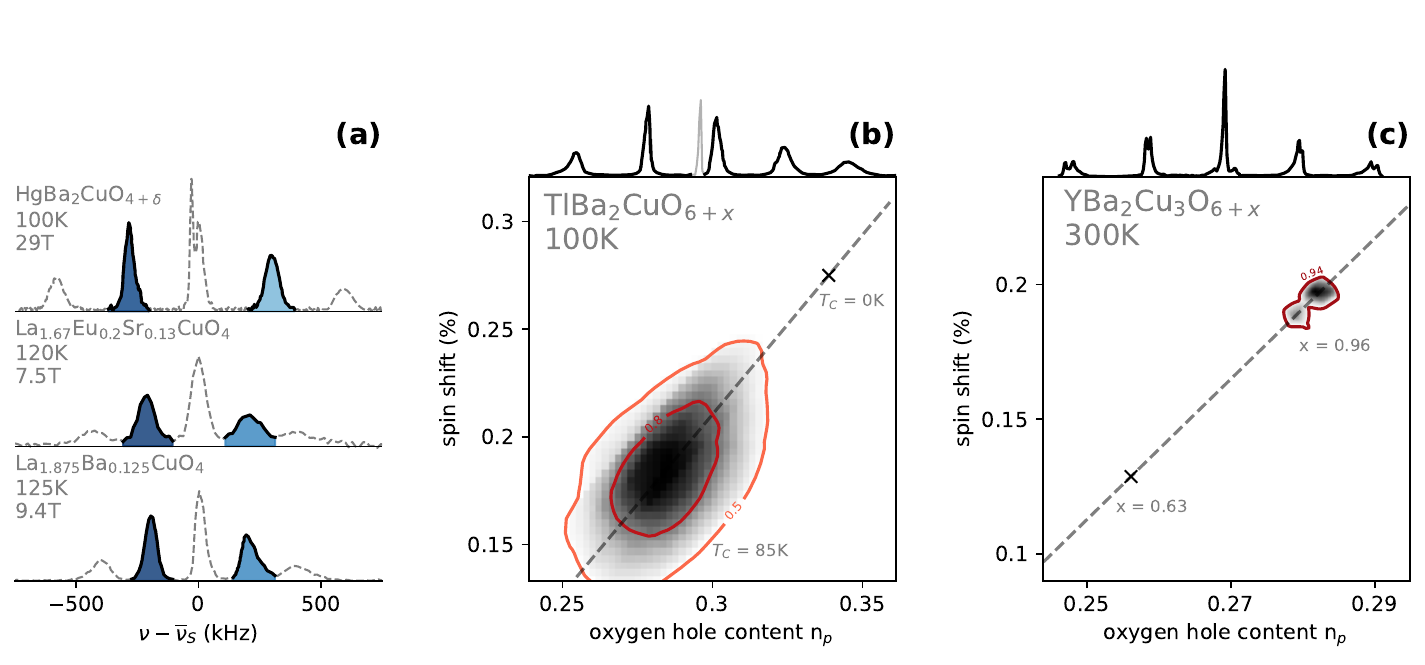}
\caption{Planar $^{17}$O NMR spectra of various cuprates show that the hole variations are very similar,  while splittings can be different. (a) Examples from the literature \cite{Grafe2006,Grafe2010,Lee2017}, where the asymmetry of the lineshape is clearly visible. (b) The probability ellipse of optimally doped TlBa$_2$CuO$_{6+x}$ \cite{Kambe1993} is aligned towards the data of the overdoped sample \cite{Kambe1993a}. This is similar to the observations in \lsco. (c) The two peaks in optimally doped \ybco \cite{Haase2003} show the same behavior as TlBa$_2$CuO$_{6+x}$ and \lsco and point in the direction of the underdoped sample \cite{Takigawa1991}. Grey lines are guides to the eye.}
\label{fig:fig4}
\end{figure}

The splittings, $\erww{\nu_\mathrm{Q}}$, are largely temperature independent, but increase with doping $x$ as holes enter the $2p_\sigma$ orbitals (concomitant changes in the Cu hole content do not affect the planar O splitting) \cite{Jurkutat2014}. However, the total splitting $\erww{\nu_\mathrm{Q}}$ has a strong family dependence in the sense that the chemistry of the family determines how the inherent Cu$^{2+}$ hole is shared between planar Cu and O \cite{Jurkutat2014}. Interestingly, it was found that the maximum $T_\mathrm{c}$ of a family is apparently proportional to the total planar O hole content, $\erww{n_\mathrm{p}}$, at optimal doping, including that is inherent to the parent compound \cite{Rybicki2016}. The differences in the total oxygen hole content are apparent in Fig.~\ref{fig:fig4}. The shifts, of course, follow the doped hole content and, as such, they are quite independent of the specific cuprate. The interlayer chemistry sets the sharing of the inherent hole between Cu and O in the parent material, as well as how doped holes enter the plane and split between Cu and O \cite{Jurkutat2014,Jurkutat2023}.

%%%%%%%%%%%%%%%%%%%%%%%%%%%%%%%%%%%%%%%%%%%%%%%%
%%%%%%%%%%%%%%%%%%%%%%%%%%%%%%%%%%%%%%%%%%%%%%%%%%

A summary of all planar $^{17}$O NMR shifts was published recently \cite{Nachtigal2020}. At high doping levels and above $T_\mathrm{c}$, the shifts are nearly temperature independent, as expected for a normal metal (see further below about the notion of metal used here). Even the Korringa relation is approximately obeyed. At $T_\mathrm{c}$ the shift drops and disappears rapidly at low temperatures (spin-singlet pairing).
As doping decreases, the rapid change at $T_\mathrm{c}$ disappears gradually and the temperature-independent part of the shift at high temperatures develops a drop in magnitude, i.e.\@ a high-temperature offset. This is the newly discovered, universal pseudogap behavior \cite{Nachtigal2020}. It follows from a temperature independent gap, $\Delta_\mathrm{PG}$, in a metallic density of states (DOS) so that excitations across the gap are gradually suppressed by the Fermi function as the temperature is lowered, cf.\@ inset to Fig.~\ref{fig:fig1}(a). In other words, even for a small pseudogap, the missing states  at high temperature are noticed in shift and relaxation.
The behavior near $T_\mathrm{c}$ is not clear as it appears to be obscured by the pseudogap (if the concept still holds at these temperatures).\par\medskip

Of particular importance is the empirical relation between $\erww{\nu_\mathrm{Q}}$ and $\erww{\nu_\mathrm{S}}$ that we here consider at room temperature. The fact that the left lower satellite in Fig.~\ref{fig:fig1}(b) is doping independent says that at our field strength and room temperature, the \emph{difference} between shift and splitting is doping independent (we verified with field dependent measurements that $\erww{\nu_\mathrm{S}}$ is indeed proportional to the field). This peculiar relationship was found for powder samples of \lsco and \ybco, long ago \cite{Haase2002}. Today we know that it is the doped hole content $\Delta \erww{n_\mathrm{p}}$ that sets the shift.\par\medskip

The correlated lineshapes appear since there is a spatial correlation between $n_\mathrm{p}$ and $\erww{S_z}$, the isotropic spin along the quantization axis at planar O. Unfortunately, we have no a priori knowledge of any correlation length, neither for the spin polarization nor for the hole density. Since NMR is a static, local probe, there can be differences between NMR and scattering methods, so we rather focus on NMR related evidence. We know that $n_\mathrm{p}^j$ carries atomic scale resolution (oxygen atom $j$), however, a particular nucleus may experience an $\erww{S_z}$ that may stem from a local electronic spin, as well as that from many delocalized electronic spins as in metals. However, given the observed correlation between charge and spin, one would assume that both share a similar lengthscale. Perhaps the uncorrelated broadening could, for instance, be a result of the commensurate mismatch of the wavelengths or variations of the amplitudes. 

We do know that this correlation is very similar for all cuprates, where data are available, so it must be a property of the planar quantum matter. Of course, a mere spatial doping variation would explain our observation. Then, for \lsco we would conclude that it is the Sr dopant that induces such a doping variation, and this was indeed concluded from the study of the magnetic linewidths and relaxation alone \cite{Singer2005}. In other materials, a  mismatch of the local energy density (from impurities) might simply be behind such spatial doping variations that trigger $\Delta n_\mathrm{p}$ and $\Delta \erww{S_z}$.

As a matter of fact, additional linewidths are largely absent for stoichiometric materials such as \ybcoE, or \ybcoF, in particular also for planar Cu. However, the planar O satellites show even in these materials a splitting similar to the linewidths of \lsco. It was therefore argued long ago \cite{Haase2003} that an intra unit-cell, largely commensurate charge density variation could be behind the NMR data (instead of ascribing it to chemical orthorhombicity). Meanwhile such modulations have been discovered with STM \cite{Fujita2014}. More recently, special NMR experiments were employed for \ybco single crystals, and orientation dependent studies showed the presence of such charge density variations \cite{Reichardt2018}. As can be seen in Fig.~\ref{fig:fig4}, the satellite splittings of YBa$_2$Cu$_3$O$_{6.96}$ fit the same correlation behavior. We thus know that $n_\mathrm{p}^j$ and $n_\mathrm{p}^{j+1}$ of neighboring oxygen atoms can be rather different.

This then demands very short length scale variations in terms of charge, i.e. even within the unit cell. As shown in Fig.~\ref{fig:fig4}, the charge distribution also fits the shift distribution for this splitting. We must conclude that $\erww{S_z}^j$ is even somewhat modulated, where $j$ again points to the atomic average.

Given the independent evidence in many publications, our observations must be the NMR realization of stripe-like correlations. The local charge $\Delta n_\mathrm{p}$ correlates with a nearly local spin $\Delta \erww{S_z}^j$. This behavior is quite general, even the depth of the modulations is not very different for all cuprates. Clearly, we only see static remnants of rapidly fluctuating stripes. However, upon pinning, the response is a correlated charge and spin density wave. Through the lens of NMR's independent $n_\mathrm{p}^j$ and $\erww{S_z}^j$, one expects circles for uncorrelated signals in a plot like the one shown in Fig.~\ref{fig:fig2}, or ellipses if correlations are present (similar to slightly disturbed Lissajous figures on an oscilloscope fed with two sinusoidal signals with fixed phase). A change in phase between the two will rotate an ellipse. We think that this may be the generic interpretation of the charge and spin correlations observed with NMR in the cuprates. It is present at all doping levels (on a local scale, at least) with a phase change at $x=0.25$, after which it continues at least until $x=0.30$. This rather universal behavior casts strong doubt on a more or less normal Fermi liquid description of overdoped \lsco, and points to the critical point as a phase slip scenario.

%%%%%%%%%%%%%%%%%%%%%%%%%%%%%%%%%%%%%%%%%%%%%
%%%%%%%%%%%%%%%%%%%%%%%%%%%%%%%%%%%%%%%%%%%%%%%%
\section{Conclusions}
%%%%%%%%%%%%%%%%%%%%%%%%%%%%%%%%%%%%%%%%%%%%%%%%%%
%%%%%%%%%%%%%%%%%%%%%%%%%%%%%%%%%%%%%%%%%%%%%%%%
We conclude than that stripe-like correlations of spin and charge are a fundamental property of the cuprates, at least in the CuO$_2$ plane. The amplitude of the charge modulation is rather doping and material independent and its wavelength is short, such that neighboring O atoms can already have different $n_\mathrm{p}$. The NMR investigation of the pressure and temperature dependence of the splittings in YBa$_2$Cu$_3$O$_{6.96}$ showed \cite{Reichardt2018} that a reversible bulk ordering of the charge variations can be achieved with respect to a particular field direction \cite{Reichardt2018}. Thus, one has to expect ordering phenomena with such pinned stripes.

At least for \lsco, at a doping level of $x=1/4$ the correlation is lost where it changes sign. This could be caused by a change in phase between both oscillations. Perhaps this is to be expected at higher doping levels. This turning point might coincide with a critical point. In any case, the findings do not agree with a simple Fermi liquid picture at high doping levels. Furthermore, the apparently nearly atomic extent of $\erww{S_z}$ raises the question of the applicability of such a picture. Apparently, a Van Hove scenario is discussed for \lsco and it remains to be seen whether this is accidentally associated with the turnover in correlation, or perhaps intimately related to it. 

For doping levels below $\sim x=0.2$, where there is a pseudogap, a natural explanation for the correlations of spin and charge could be related to the pseudogap itself: as it removes states from an otherwise universal DOS \cite{Nachtigal2020}, one would argue that an inhomogeneous pseudogap, that has been found with STM long ago \cite{Lawler2010}, may cause short range variations of $\erww{S_z}$ as well. While this is a simple conclusion, the explanation does not seem to hold at higher doping levels, and for the doping independent variations in spin and charge in general. Perhaps then one can ask whether the pseudogap itself is a consequence of passing the critical point of $x=1/4$ towards lower doping levels, which is set by particular stripe correlations. 

Clearly, a more thorough investigation of the findings is advised, including for other materials. Unfortunately, 
planar Cu data do show significant family dependences and, therefore, do not lend themselves to a simple comparison. However, Cu lineshapes are known to be very different in different materials and thus could aid understanding. Finally, a thorough comparison with theoretical models and other probes' findings will be very important as well.

{\flushleft \bf Acknowledgments}\par\medskip
JH acknowledges the friendship to the late Jan Zaanen, in particular the many discussions during hideouts at conferences. We thank Leipzig University for financial support. We thank  Jakob Nachtigal (Leipzig) for many discussions, Crina Berbecariu (Leipzig) for proofreading the manuscript.
JH thanks K. Fujita and J. Tranquada for discussions at an early date.

%\appendix
%\section{Example Appendix Section}
{\flushleft \bf CRediT authorship contribution statement}\par\medskip
{\bf D. Bandur:} Conceptualization, %Data curation
Formal analysis, Investigation, Methodology, Resources, Software, Validation, Writing - original draft, Writing - review \& editing.
{\bf A. Lee:} Conceptualization, Formal analysis, Methodology, Writing - original draft, Writing - review \& editing.
{\bf S. Tsankov:} Writing - review and editing. 
{\bf A. Erb:} Investigation, Validation, Writing - review and editing, 
{\bf J. Haase:} Conceptualization, Resources, Funding acquisition, 
 Methodology, Project administration, Supervision, Validation, Writing - review and editing, Writing - original draft, 

%%       \cite{<label>} ==> [1]
{\flushleft \bf Declaration of competing interest}\par\medskip
The authors declare that they have no known competing financial interests or personal relationships that could have appeared to influence the work reported in this paper.
%%
%Example citation, See \cite{lamport94}.

{\flushleft \bf Data availability}\par\medskip
Data will be made available on request.

{\flushleft \bf Funding information\\}
Funding of the research came from Leipzig University.\par\medskip

\bibliography{JH-CuprateJHb.bib}   % APS-like style for physics

\printindex

\par\medskip

\end{document}